\documentclass[aps,twocolumn,pra,superscriptaddress,showpacs,showkeys,floatfix]{revtex4-1}

\usepackage{amsbsy}
\usepackage[pdftex]{graphicx}
\usepackage{caption}
\usepackage{amsmath}
\usepackage{amssymb}
\usepackage{amsfonts}
\usepackage{amssymb}
\usepackage{float}
\usepackage{graphicx}
\usepackage{enumerate}
\usepackage{epstopdf}
\usepackage{braket}
\usepackage{bm}
\usepackage{color,soul}

\def\hbn{{\hfill\break\noindent}}
\begin{document}

\title{Optimal control of a probabilistic dynamic for epidemic spreading in arbitrary
complex networks}

\author{Fabrizio Angaroni}

\affiliation{Center for Nonlinear and Complex Systems,
Dipartimento di Scienza e Alta Tecnologia,
Universit\`a degli Studi dell'Insubria, via Valleggio 11, 22100 Como, Italy.}
\affiliation{Istituto Nazionale di Fisica Nucleare, Sezione di Milano,
via Celoria 16, 20133 Milano, Italy.}
\begin{abstract}
\begin{center}
PRE-PRINT-VERSION\\
\end{center}
This paper presents a discrete time probabilistic dynamic for simulating a 
contact-based epidemic spreading based on discrete time Markov chain process, 
in particular the attention is addressed to the susceptible-infectious-removed (SIR)
model and the phase diagram of such model will be presented.
Then, this report presents the
set of equations that represent the optimal control strategies, 
by the means of Pontryagin's maximum principle,
in two different cases
a vaccination policy and a combined vaccination-hospitalization policy
and show a numerical simulation, with the standard forward-backward sweep procedure,
for these equations.    
\end{abstract}

    \maketitle
    

\section{Introduction}
Modeling how diseases spread among individuals
was introduced by Kermack and McKendrick \cite{cit:SIR, book-intro, kiss-book} in 1927; they introduced
a model known as the susceptible-infectious-removed (SIR)
epidemic model, and they supposed that individuals of a population could
be divided into three non-intersecting classes: 
susceptible who are healthy but can contract the disease,
infectious individuals who have contracted the disease and can transmit it, and
removed individuals who have recovered and cannot contract the disease anymore.
If the epidemics is
assumed to be Markoffian and proceeds 
via infection of nearest neighbors, then the relations that represent
the SIR model are 
\begin{equation}
\begin{array}{c}
 S+I \to I +I, \\
 I\to R.
\label{eq:gep}
\end{array}
 \end{equation}
This kind of problem is called the general epidemic process (GEP) \cite{grassberger-1};
it is a stochastic multiparticle process that describes a vast number
of phenomena in nature; for example, a simple stochastic and
space-dependent model in regular lattice could describe
the fairy rings, chemical solitary waves \cite{solitary-waves-2, solitary-waves-1}, percolation \cite{percolation-weiss}, forest fires, and 
contact-based epidemic spreading.\\
The study complex networks has provided a better tool study
 the GEP in complicated topology that describe the biological system:
for example, scale-free networks
can represent sexual contacts, the Internet, and many other
social, technological, and biological networks \cite{cite-network-2, complex-net-book, cite-network-1}.\\
One of the most important goals of these studies \cite{control-rev} is to find
an optimal feedback control strategy to minimize the epidemics,
and optimality in this case means balancing some
control costs against performance. For example, the optimal solution could provide 
the best vaccination or hospitalization strategy to minimize the cost
of an epidemic \cite{opt-control-polonica}.\\
Most of the studies on this problem use the
mean-field (MF) approximation \cite{control-rev, complex-net-book}, and these kinds of studies have 
evaluated the macroscopic feature of the system-like critical behavior and phase transition 
  \cite{phase-diagram, critical-lattice}; however, MF is not designed to provide information about individual nodes.
To obtain information at
the individual level (microscopic) of description, it is necessary to use
Monte Carlo (MC) simulations. The main problem of MC models is the computational
effort needed to evaluate the expectation values of microscopic quantities.
Another point against Monte Carlo approach 
is that the optimal solution must be found through
a random search
algorithm, and the computational effort could be cumbersome
due to the dimension of the space of possible 
configurations of a complex network.\\
In this paper, the microscopic Markov-chain approach (MMCA)
\cite{sergi} for the SIR model
is presented, and the vaccinated class to the model is then added for 
the optimal control system for this model in two different cases: 
the first case considers only a 
vaccination of susceptible policy, and the second case only considers two possible
control strategies: the vaccination of the susceptible and the hospitalization of the infected.
Some numerical examples are also presented.\\
\section{The model}
\begin{figure}
\centering
\includegraphics[width=7cm ]{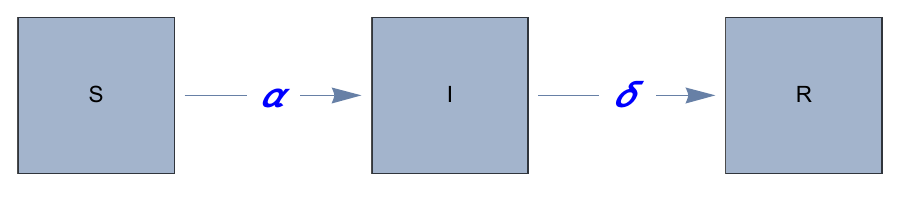}\\
\includegraphics[width=7cm ]{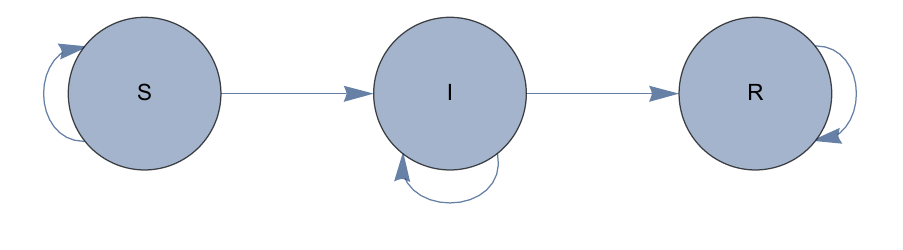}
\caption{SIR.}
\label{fig:SIR}
\end{figure}

Consider the dynamics of a SIR
epidemic process over a complex network
composed by $N$ nodes and
the example of compartments diagram
that represent the SIR model shown in Figure \ref{fig:SIR} (top).
The topology of the contacts of the network is completely determined by its adjacency 
matrix \cite{complex-net-book} $\mathbb{A}$.
Without loss of generality, $\mathbb{A}$ is assumed to be symmetric and its elements
$a_{i,j}$ are assumed to be only two possible values: $0$ and $1$. This hypothesis is not 
necessary, and in general cases where the strength of the link is considered can be treated 
in the same manner of this work.
Following the construction of Johnson \cite{Johnson} for a stochastic dynamic, 
a vector space $\mathbb{V}_k \quad k=1,\dots N$ is associated with every node; 
and  defining a basis, a non-natural isomorphism $\mathbb{V}_k\backsimeq\mathbb{R}^{d_k}$ is built,
and where $d_k$ are the possible configurations of $k$-th node,
which $d_k=d$  is supposed $\forall k$. 
Then, a vector that represents its state is attached to each node. 
For example, in the SIR model, the following state is obtained for every node:
\begin{equation}
 \overline{v}_k(t_i)=
 \begin{bmatrix}
  S_k(t_i)\\
  I_k(t_i)\\
 R_k(t_i)\\ 
  \end{bmatrix}.
 \end{equation}
Components of vectors
$A_k (t_i)$ for $A=S,I,R$ are treated as the probability that $k$-th node is
in the configuration $A$
at time $t_i$. This hypothesis implies that every
$ \overline{v}_k(t_i)$ is normalized by the means of the 1-norm, or
\begin{equation}
 S_k(t_i)+I_k(t_i)+R_k(t_i)=1.
\end{equation}
In addition, 
the systems 
is represented by a state $\Lambda (t_i)\in \otimes_{k=1}^N \mathbb{V}_k$.\\
$\alpha$ is defined as the probability for the unit of time that an infectious individual 
spreads the disease to one of its neighbors and $\delta$ as the probability for unit of time 
that an infected node is removed.
Then, the probability of node $k$ not being infected by any neighbor is the following:
\begin{equation}
 P_{0,k}(t_i):=\prod_{j=1}^N (1-a_{k,j}\alpha I_{j}(t_k)),
\end{equation}
and the discrete-time equation that describes the dynamics are given by
\begin{equation*}
 S_{k}(t_{i+1})=S_{k}(t_{i})-(1-P_{0k}(t_{i})S_{k}(t_{i}),
\end{equation*}
\begin{equation}
 I_{k}(t_{i+1})=I_k(t_i)+(1-P_{0,k}(t_i))S_{k}(t_{i})-\delta I_k(t_i)
 \label{eq:dyn-SIR}
\end{equation}
\begin{equation*}
  R_{k}(t_{i+1})=1-S_{k}(t_{i+1})-I_k(t_{i+1})
\end{equation*}
This kind of formalism in \cite{sergi} is called MMCA, 
since it is equivalent to associate a discrete 
time Markov chain for every node of the network; 
for example, see Figure \ref{fig:SIR} (bottom) for the microscopic Markov chain of the SIR model.
The solution is easily obtained if initial conditions are given,
and in \cite{sergi}, it is proven that this formalism can outperform
heterogeneous MF (HMF) and MC approaches.\\
Equations \ref{eq:dyn-SIR} can always be solved if an initial condition is given, and
it is possible to accommodate initial conditions with some grade of uncertainty, 
which is common dealing with real data.\\ 
It is known that the SIR model
has two different phases \cite{phase-diagram, grassberger-1, grassberger-2, percolation-weiss} (equilibrium states) for $t_{fin}\to \infty$ , 
one where  all the susceptible become
infected or removed another where some susceptible
manage to survive without  contracting the disease.

\section{Single control}
\begin{figure}
\centering
\includegraphics[width=3cm ]{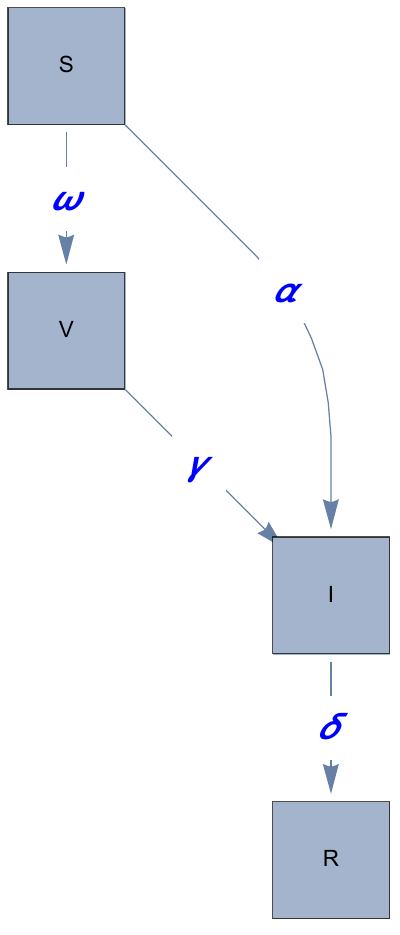}
\includegraphics[width=4cm ]{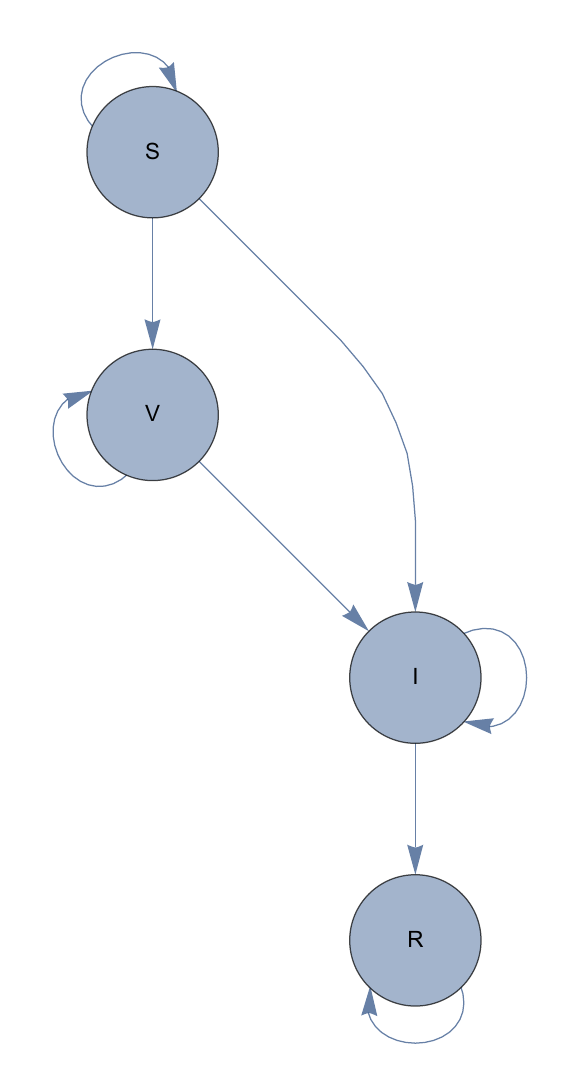}
\caption{At left there is the flowchart associated to the SIVR with one control,
at right there is the MMC for this problem}
\label{fig:SIRV}
\end{figure}

This section shows the construction of an optimal control system for the discrete time
process presented in \cite{chapman-book, opt-control-ex-2, sethi-book}. 
In addition, there is a policy maker that is supposed to be able to determine when it is necessary
to vaccinate a node. To simulate this procedure, the vaccinated class was added
to the system, and the model in this case is called susceptible-infectious-removed -vaccinated (SIRV) \cite{book-intro};
Now, the possible micro states of a node are $d=4$.
The set of admissible controls
(possible actions of the policy makers) is assumed to be provided by the following:
\begin{equation}
  \begin{array}{c}
  \mathbb{U}=\{ \omega_1(t_1),\dots\omega_N(t_{fin})
 |\forall j=1,\dots N, \forall l=0,\dots t_{fin},\\
 0\leq \omega_{j}(t_{l}) \leq 1 \};\\
  \end{array}
 \end{equation}

where $\omega_k(t_i)$ is the probability at time $i$ that the $k$-th
susceptible node becomes vaccinated; here some grade of failure of the therapy is assumed 
in fact it is possible that a vaccinated becomes infectious with probability 
$\gamma<\alpha$, this represents a non-efficacious vaccination \cite{book-intro} 
or a development of the resistance 
to the disease \cite{vax-plant}.
The probability of the vaccinated
node $k$ not being infected by any neighbor is defined as the following:
\begin{equation}
 P_{0,k,V}(t_i):=\prod_{j=1}^N (1-a_{k,j}\gamma I_{j}(t_k)),
\end{equation}
where $V_{k}(t_i)$ is the probability that the $k$-th node is vaccinated at time $t_i$.\\
Then, the dynamics of the probabilities is driven by the following $4N$
finite difference time equations:
\begin{equation*}
\begin{array}{c}
 S_{k}(t_{i+1})=S_k(t_i)-(1-P_{0,k}(t_i))S_{k}(t_i)(1-\omega_k(t_i))\\
 -P_{0,k}(t_i)\omega_k(t_i)S_k(t_i),\\
 \end{array}
  \end{equation*}

\begin{equation}
\begin{array}{c}
 I_{k}(t_{i+1})=I_k(t_i)+(1-P_{0,k}(t_i))S_{k}(t_i)(1-\omega_k(t_i))\\
+(1-P_{0,k,V}(t_i))V_{k}(t_i)-\delta I_k(t_i)\\
\end{array}
 \label{eq:SIRV}
\end{equation}
\begin{equation*}
\begin{array}{c}
 V_{k}(t_{i+1})=V_k(t_{i})-(1-P_{0,k,V}(t_i))V_{k}(t_i)\\
+P_{0,k}(t_i)S_k(t_{i})\omega_k(t_i),\\
\end{array}
\end{equation*}
\hbn
\begin{equation*}
 R_{k}(t_{i+1})=1-S_k(t_{i+1})-I_k(t_{k+1})-V_k(t_{k+1}).
\end{equation*}
The last equation is obtained from the normalization condition.
The compartment diagram for this model shown in Figure \ref{fig:SIRV} 
(left) instead of in Figure
\ref{fig:SIRV} (right) shows the microscopic Markov chain associated
 to every node of the complex network.\\
To proceed further, the following $3N$ functions must be defined:
\begin{equation*}
\begin{array}{c}
 g_{k,S}(t_i)= S_k(t_{i+1})-S_k(t_i)=\\
=-S_k(t_i)(1-P_{0,k})(1-\omega_k(t_i))S_k(t_i)\\
-P_{0,k}\omega_k(t_i)S_k(t_i),\\
\end{array}
\end{equation*}

\begin{equation}
\begin{array}{c}
 g_{k,I}(t_i)= I_k(t_{i+1})-I_k(t_i)=\\
=(1-P_{0,k}(t_i))(1-\omega_k(t_i))S_k(t_i)\\
-\delta I_k(t_i)+(1-P_{0,k,V}(t_i))V_k(t_i),\\
\end{array}
\end{equation}

\begin{equation*}
\begin{array}{c}
 g_{k,V}(t_i)= V_k(t_{i+1})-V_k(t_i)=\\
=-(1-P_{k,0,V}(t_i))V_k(t_i)+P_{0,k}\omega_k(t_i)S_k(t_i).\\
\end{array}
\end{equation*}

In addition, the cost must be defined:
\begin{equation}
 J(\Lambda_{fin},t_{fin}):=
 \sum_{k=1}^N\left[\phi_k(t_{fin})+\sum_{j=0}^{t_{fin}-1}f_{j,k}(\Lambda_j,t_j)\right],
\end{equation}
the first term represents
a payoff given by the final state, while the second term represents a 
cumulative cost. Without loss of generality, 
the maximum is given by the following:
\begin{equation}
  J(\Lambda_{fin},t_{fin})=-\sum_{k=1}^N\left[\sum_{j=0}^{t_{fin}-1} C_1I_k(t_j)
  +\frac{C_2}{2}\omega^2_k(t_j)  \right].
  \label{eq:cost}
\end{equation}
This cost simulates when there is a cost associated to 
the vaccination and a cost associated to the infectious individuals,
which, for example, can represent the loss of work days.\\
This definition fixes 
the behavior of the control, and it could be defined in a different 
manner to change the behavior of the policy maker, 
for example, see \cite{opt-control-polonica}.
However, the only  
request is that the cost 
is quadratic in the control (i.e. $\omega_k(t_i)$), and the same procedure of this 
work could be used to consider more realistic costs.
By choosing a finite horizon optimal control,
adding a time discount to the cost creates
an infinite horizon
control. However, the first term of \eqref{eq:cost} gives a cost that increases 
with time due to the intrinsic dynamics of the system,
so it is not necessary to use a time discount function.
In addition,
if cost depends onto $R_k(t_i)$, then it is not possible to determine $R_k(t_i)$
from the normalization condition.\\
Then, to build an optimal control system, a Hamiltonian must be defined:
\begin{equation}
 H(t_i,\Lambda_i,\omega_i)=H_i=\sum_{k=1}^N [ f_{k,i} +\lambda_{k,S}(t_{i+1})g_{k,S}(t_{i})
 \label{eq:ham}
 \end{equation}
\begin{equation} + \lambda_{k,I}(t_{i+1})g_{k,I}(t_{i})
 +\lambda_{k,V}(t_{i+1})g_{k,V}(t_{i})],
\end{equation}
where $\lambda_{k,A}(t_i)\quad A=S,I,V$ are the adjoint functions. 
First, the transversality 
conditions are given by the following:
\begin{equation}
 \lambda_{k,A}(t_{fin})=\frac{\partial \phi_k(t_{fin})}{\partial A_k(t_{fin})}=0.
 \label{eq:trasv}
\end{equation}
Then, from the definition \eqref{eq:ham},
the adjoint functions for $t\neq t_{fin}$ are given by the following
\begin{equation}
 \lambda_{k,S}(t_i)=\frac{\partial H_i}{\partial S_{k}(t_i)},
\end{equation}
\begin{equation}
 \lambda_{k,I}(t_i)=\frac{\partial H_i}{\partial I_{k}(t_i)},
\end{equation}
\begin{equation}
 \lambda_{k,V}(t_i)=\frac{\partial H_i}{\partial V_{k}(t_i)}.
\end{equation}
Thus, the following recurrence relations are obtained for the adjoint functions by solving the previous equations:
\begin{equation*}
\begin{array}{c}
  \lambda_{k,S}(t_i)=\lambda_{k,S}(t_{i+1})[-(1-P_{0,k}(t_i))(1-\omega_k(t_i))\\
  -P_{0,k}(t_i)\omega_k(t_i)]+\\
\lambda_{k,I}(t_{i+1})[(1-P_{0,k}(t_i))(1-\omega_k(t_i))\\
+\lambda_{k,V}(t_{i+1}P_{0,k}(t_i)\omega_k(t_i)],\\
\end{array}
 \end{equation*} 

\begin{equation}
\begin{array}{c}
\lambda_{k,I}(t_i) = C_1- \delta \lambda_{kI}(t_{i+1})+\\
 \sum_{j=1\quad j\neq k}^N\{  +\lambda_{j,S}(t_{i+1}) \\
{[(1-\omega_j(t_i)S_k(t_i)h_{k,j}(t_i)}   \\
 -\omega_j(t_i)S_j(t_i) h_{k,j}(t_i) ] \\
 -\lambda_{j,I} (t_{i+1})[ (1-\omega_j(t_i)S_j(t_i)h_{k,j}(t_i))\\
 +V_j(t_i)hv_{k,j}(t_i)  ]) +\\
\lambda_{j,V}(t_{i+1})[V_j(t_i)hv_{k,j}(t_i)\\
+S_j(t_i)\omega_j(t_i)h_{k,j}(t_i)]\},
 \end{array}
 \label{eq:adjoint-SIRV}
 \end{equation}

\begin{equation*}
\begin{array}{c}
   \lambda_{k,V}(t_i)=\lambda_{k,I}(t_{i+1})(1-P_{0,k,V}(t_i))\\
   +\lambda_{k,V}(t_{i+1})(P_{0,k,V}(t_i)-1)
   \end{array}
\end{equation*}

where $h_{k,j}(t_i)$ and $hv_{k,j}(t_i)$ are given by the following:
\begin{equation}
 h_{k,j}(t_i)=-a_{j,k}\alpha\prod_{l=1,l\neq k}^N(1-a_{j,l}\alpha I_l(t_i)),
 \label{eq:defh}
\end{equation}
\begin{equation}
 hv_{k,j}(t_i)=-a_{j,k}\gamma\prod_{l=1,l\neq k}^N(1-a_{j,l}\gamma I_l(t_i)).
 \label{eq:defhv}
\end{equation}
Eventually, by applying the extension of Pontryagin’s
maximum principle for discrete systems \cite{pontryagin}, 
the characterization of the optimal control is obtained,
imposing $\frac{\partial H_i}{\partial \omega_k (t_i)}=0$:
\begin{equation}
 \omega^*_k(t_i)=\min\{\max\{ 0,K \},1 \},
\end{equation}
with
\begin{equation}
\begin{array}{c}
 K=\frac{1}{C_2}\{  \lambda_{k,S}(t_i+1)[(1-P_{0,k}(t_i))S_k(t_i)\\
 -P_{0,k}(t_i)S_k(t_i)]+ \\
 +\lambda_{k,I}(t_i)[-(1-P_{0,k}(t_i))S_k(t_i)]+\lambda_{k,V}(t_i)P_{0,k}S_k(t_i)\}.
 \label{eq:opt-single}
 \end{array}
 \end{equation}
The optimality system in the state Equations 
\eqref{eq:SIRV} with initial conditions and adjoint Equations \eqref{eq:adjoint-SIRV} with the final time conditions
(transversality conditions) and with the
characterization of the optimal control \eqref{eq:opt-single}.
Some numerical examples are also shown in section \ref{sec:numerical}
\section{Double Control}
\begin{figure}
\centering
\includegraphics[width=6cm ]{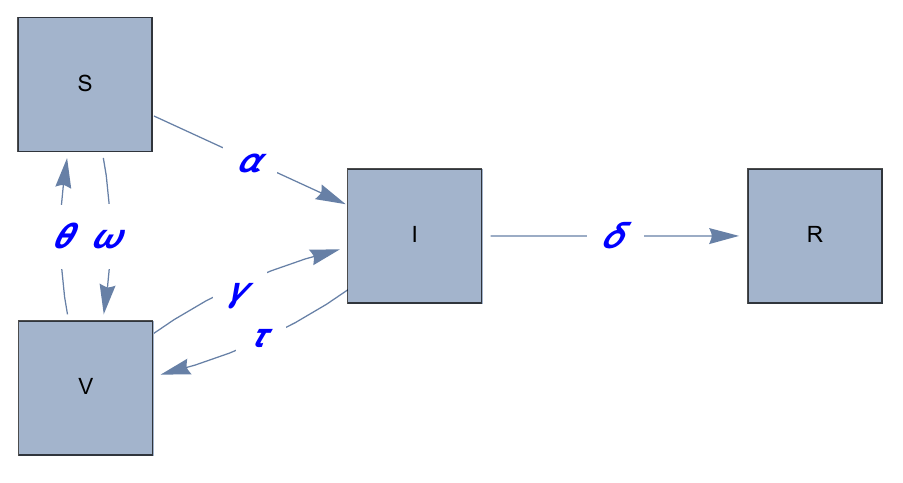}\\
\includegraphics[width=6cm ]{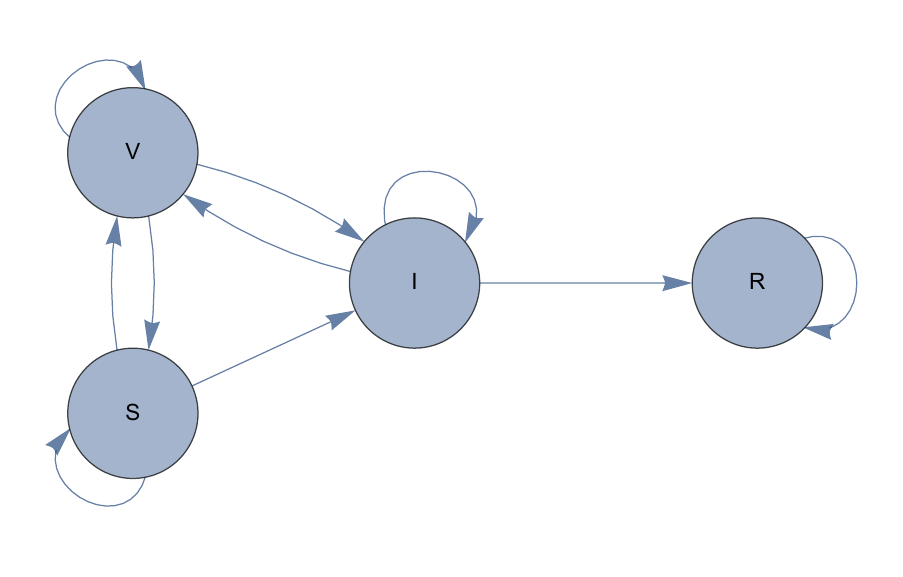}
\caption{At left there is the compartment diagram associated to the SIVR with double control, at bottom there is the MMC for this problem.}
\label{fig:double}
\end{figure}

Now,  a more general case of a contact epidemic it  is consider supposing that there is a natural
time decay of the protection given by the
vaccination,
there is a probability ($\theta$) for a unit of time
that a vaccinated becomes susceptible. 
In addition, there are the possibility that the policy maker could cure an infected,
for example, the hospitalization of a person with an infectious disease, 
this action is represented by a transition from infected to vaccinated.
Thus, the probability for a unit of time of this event is given by the new set of control
$\tau_k(t_i)$, there are $2N$ controls,
and the set of possible actions of the policy maker
is given by the following equation:
\begin{equation}
 \mathbb{U}=\{ \omega_1(t_1),\dots\omega_N(t_{fin}), \tau_1(t_1),\dots \tau_N(t_{fin})
 \end{equation}
$$
 |\forall j=1,\dots N, \forall l=0,\dots t_{fin}, 0\leq \omega_{j}(t_{l}) \leq 1 ,0\leq \tau_{j}(t_{l}) \leq 1 \}.
$$
The compartment diagram for such system is presented in Figure \ref{fig:double} (top).
Given the initial conditions, 
the discrete time dynamics of this system is completely
specified by the following $4N$ equations :
\begin{equation*}
\begin{array}{c}
 S_{k}(t_{i+1})=S_k(t_i)-(1-P_{0,k}(t_i))\\
 (1-\omega_k(t_i))S_k(t_i)-P_{0,k}\omega_k(t_i)S_k(t_i)+\\
 \theta V_k(t_i)P_{0,V,k}(t_i),
 \end{array}
 \end{equation*}

\begin{equation}
\begin{array}{c}
I_{k}(t_{i+1})=I_{k}(t_i)+(1-P_{0,k}(t_i))(1-\omega_k(t_i)S_{k}(t_i)\\
+(1-P_{0,V,k}(t_i))(1-\theta)V_k(t_i)-(1-\delta)\tau_k(t_i)I_k(t_i)\\
-\delta I_k(t_i)(1-\tau_k(t_i)),\\
\end{array}
\label{eq:dynamics:double} 
\end{equation}
\begin{equation*}
\begin{array}{c}
V_k(t_{i+1})  =V_k(t_i)+P_{0,k}(t_i)\omega_k(t_i)S_k(t_i) \\
+(1-\delta)\tau_k(t_i)I_k(t_i)\\
-(1-P_{0,V,k}(t_i))(1-\theta)V_k(t_i),
\end{array}
\end{equation*}
\begin{equation*}
 R_k(t_i)=1-S_k(t_i)-I_k(t_i)-R_k(t_i).
\end{equation*}
Again, these equations completely specify the time dynamics of the 
probability if initial (probabilistic) conditions are given.
In Figure \ref{fig:double} (bottom), the microscopic Markov chain associated
to Equations \eqref{eq:dynamics:double} is presented.\\
In this case, the cost that is maximized in the following sections is the following:
\begin{equation}
  J(\Lambda_{fin},t_{fin})=-\sum_{k=1}^N\left[\sum_{j=0}^{t_{fin}-1} C_1I_k(t_j)
  +\frac{C_2}{2}\omega^2_k(t_j) + \frac{C_3}{2}\tau^2_k(t_j)\right].
  \label{eq:cost-double}
\end{equation}
In this case, a term that can mimic hospitalization and drug
treatment costs is used.\\
As performed in the previous section, a Hamiltonian can be defined to obtain the 
recurrence relations for the adjoint functions:
\begin{equation*}
\begin{array}{c}
 \lambda_{k,S}(t_i)=\lambda_{k,S}(t_{i+1})\\
 {[-(1-P_{0,k}(t_i)(1-\omega_{0,k}(t_i))-P_{0,k}(t_i)\omega_k(t_i)]}\\
+\lambda_{k,I}(t_{i+1})(1-P_{0,k}(t_i)(1-\omega_k(t_i))+\lambda_{k,V}(t_{i+1})P_{0,k}(t_i),\\
\end{array}
\end{equation*}
\begin{equation}
\begin{array}{c}
 \lambda_{k,I}(t_i)=C_2+\lambda_{k,V}(t_{i+1})(1-\delta)\tau_k(t_i)\\
 -\lambda_{k,I}(t_{i+1})(1-\delta)\tau_k(t_i)\\
+\sum_{j=1,j\neq k}^N\{\lambda_{j,s}(t_{i+1})[S_j(t_i)(1-\omega_j(t_i)h_{j,k}(t_i)\\
-\omega_j(t_i)S_j(t_i)h_{j,k}(t_i)+\theta V_j(t_i)hv_{j,k}(t_i)]\\
+\lambda_{j,I}(t_{i+1})[-(1-\omega_j(t_i))S_j(t_i)h_{j,k}(t_i)\\
-(1-\theta)V_j(t_i)hv_{j,k}(t_i)]\\
+\lambda_{j,V}(t_{i+1})[(1-\theta)V_j(t_i)h_{j,k}(t_i)+\omega_k(t_i)S_k(t_i)h_{j,k}(t_i)\\
-\theta V_j(t_i)hv_{j,k}(t_i)]\},\\
\end{array}
 \label{eq:adjoint-double} 
\end{equation}
\begin{equation}
\begin{array}{c}
 \lambda_{k,V}(t_i)=\lambda_{k,S}(t_{i+1})\theta P_{0,V,k}(t_i)\\
+\lambda_{k,I}(t_{i+1})(1-P_{0,V,k}(t_i))(1-\theta)\\
+\lambda_{k,V}(t_{i+1})[-(1-P_{0,V,k}(t_i))(1-\theta)-\theta P_{0,V,k}(t_i)],\\
\end{array}
\end{equation}

where the functions $h_{j,k}(t_i)$ and $hv_{j,k}(t_i)$ are defined in Equations \eqref{eq:defh} and
\eqref{eq:defhv}.
The transversality conditions are the same as previously given (see Equations \eqref{eq:trasv})
Again, by imposing the following condition, 
  \begin{equation}
   \frac{\partial H_i}{\partial \omega^*_k(t_i)}=0,\qquad \frac{\partial H_i}{\partial \tau^*_k(t_i)}=0
  \end{equation},
the following $2N$ equations that characterize the optimal control are obtained:
\begin{equation}
 \omega^*_k(t_i)=\min\{\max\{ 0,K \},1 \},
\end{equation}
with
\begin{equation}
\begin{array}{c}
 K=\frac{1}{C_2}\{  \lambda_{k,S}(t_i+1)[(1-P_{0,k}(t_i))S_k(t_i)-P_{0,k}(t_i)S_k(t_i)]+ \\
 \lambda_{k,I}(t_i)[-(1-P_{0,k}(t_i))S_k(t_i)]+\lambda_{k,V}(t_i)P_{0,k}S_k(t_i)\},\\
  \label{eq:opt-double-1}
  \end{array}
 \end{equation}
\begin{equation}
 \tau^*_k(t_i)=\min\{\max\{ 0,M \},1 \},
\end{equation}
with
\begin{equation}
\begin{array}{c}
 M=\frac{1}{C_3}\{  \lambda_{k,I}(t_i+1)[-(1-\delta)I_k(t_i)+\delta I_k(t_i)]+\\
 \lambda_{k,V}(t_i)[(1-\delta)I_k(t_i)].\\
  \label{eq:opt-double-2}
  \end{array}
 \end{equation}
The optimality system consists of state Equations \eqref{eq:dynamics:double}
with initial conditions 
and adjoint Equations \eqref{eq:adjoint-double} with the final time conditions
(transversality conditions) and with the
characterizations of the optimal control \eqref{eq:opt-double-1} and \eqref{eq:opt-double-2}.

\section{numerical examples}
\label{sec:numerical}
\begin{figure}
\centering
\includegraphics[width=6cm ]{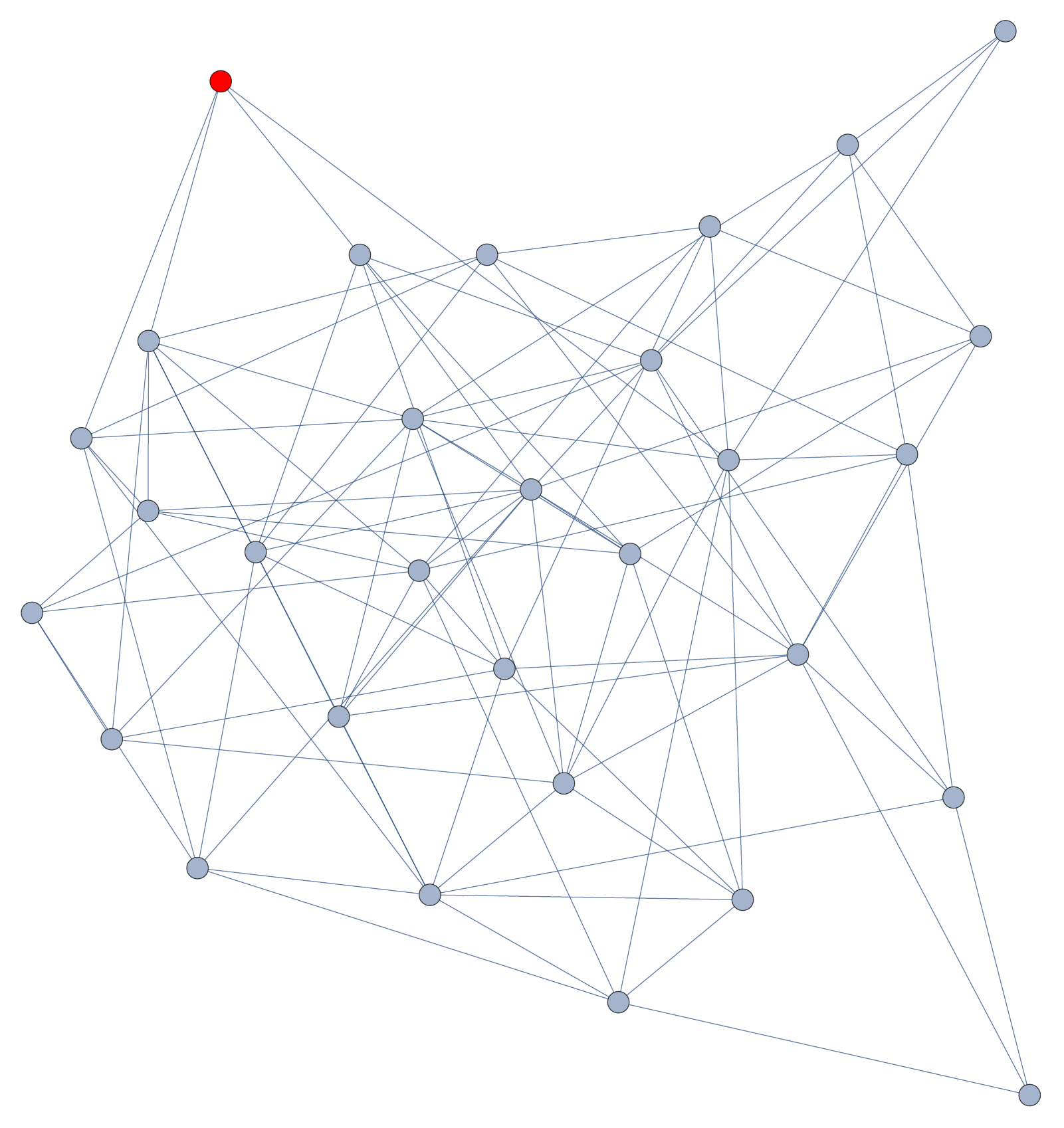}
\caption{Reference complex network used for numerical simulation. In red the node who starts infectious.}
\label{fig:complex-net}
\end{figure}

\begin{figure}
\centering
\includegraphics[width=7cm ]{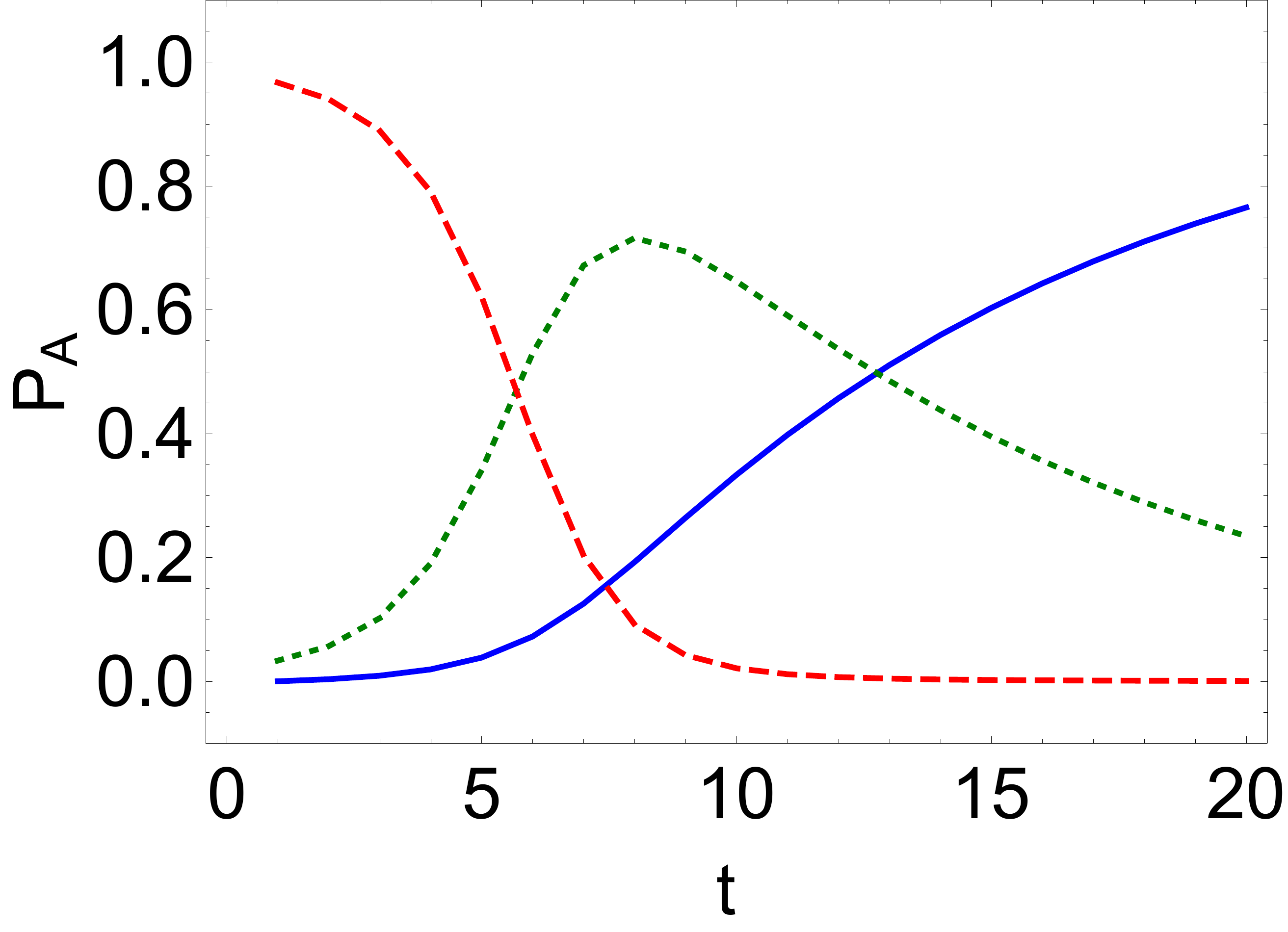}
\caption{Time dynamics of the probabilities in Equation \eqref{eq:prob} for the SIR dynamics with $\alpha=0.2$ $\delta=0.1$. The Red dashed line represents $P_S(t_i)$, 
the green dotted line represents $P_I(t_i)$, and the blue solid line represents $P_R(t_i)$. It is possible
to notice that with these values of $\alpha$ and $\delta$, susceptible-free equilibrium is reached for a sufficient time.}
\label{fig:SIR-dyn}
\end{figure}

\begin{figure}
\centering
\includegraphics[width=6cm ]{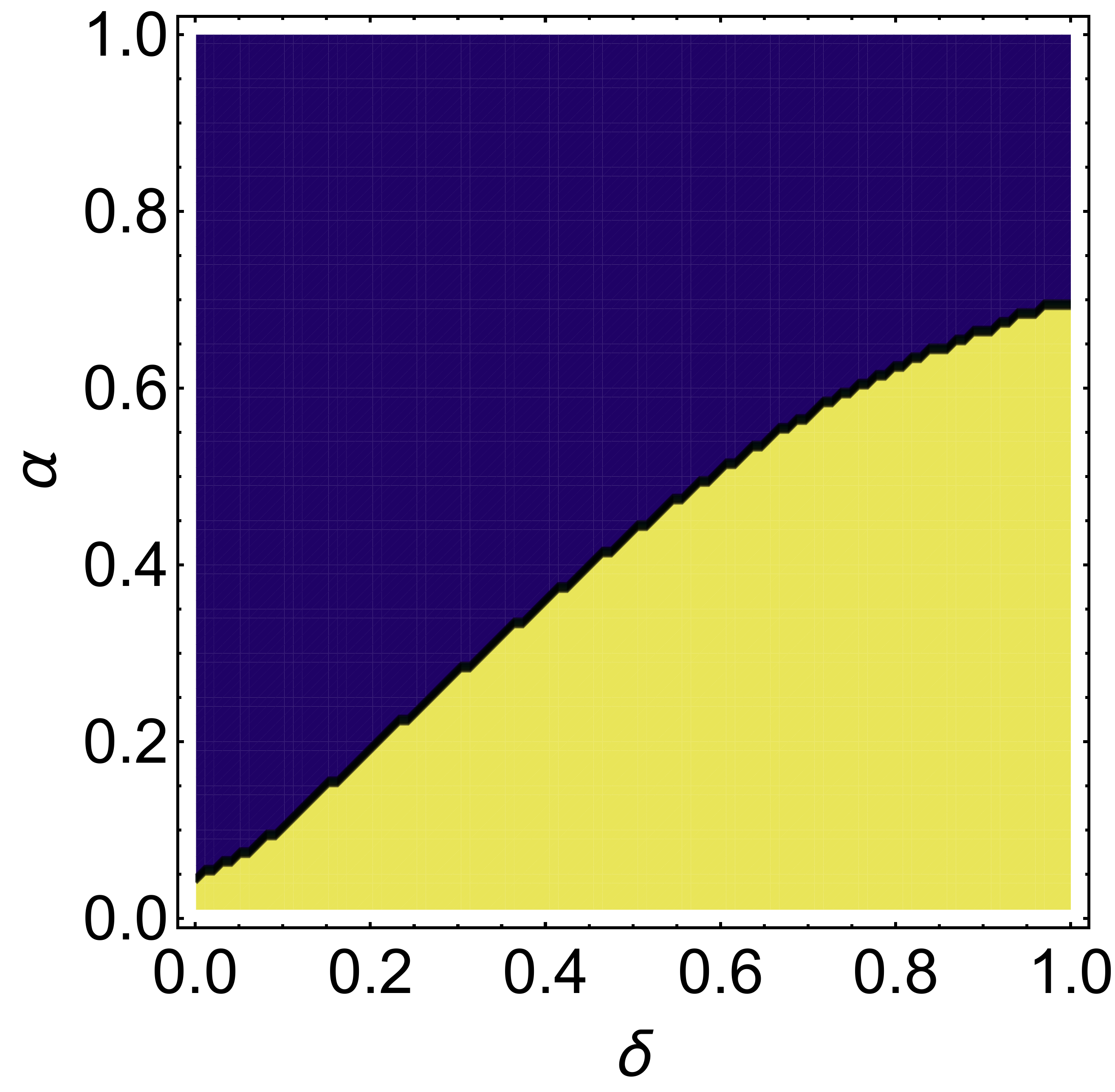}
\caption{Phase diagram of the SIR model in the reference network,  in yellow the points in the parameters space  of
$\alpha$ and $\delta$ where there is a non negligible (greater than $10^{-5}$)
probability of the survival of a healthy susceptible, in blue the points of susceptible-free equilibrium.}
\label{fig:phase}
\end{figure}
\begin{figure}
\centering
\includegraphics[width=3.5cm ]{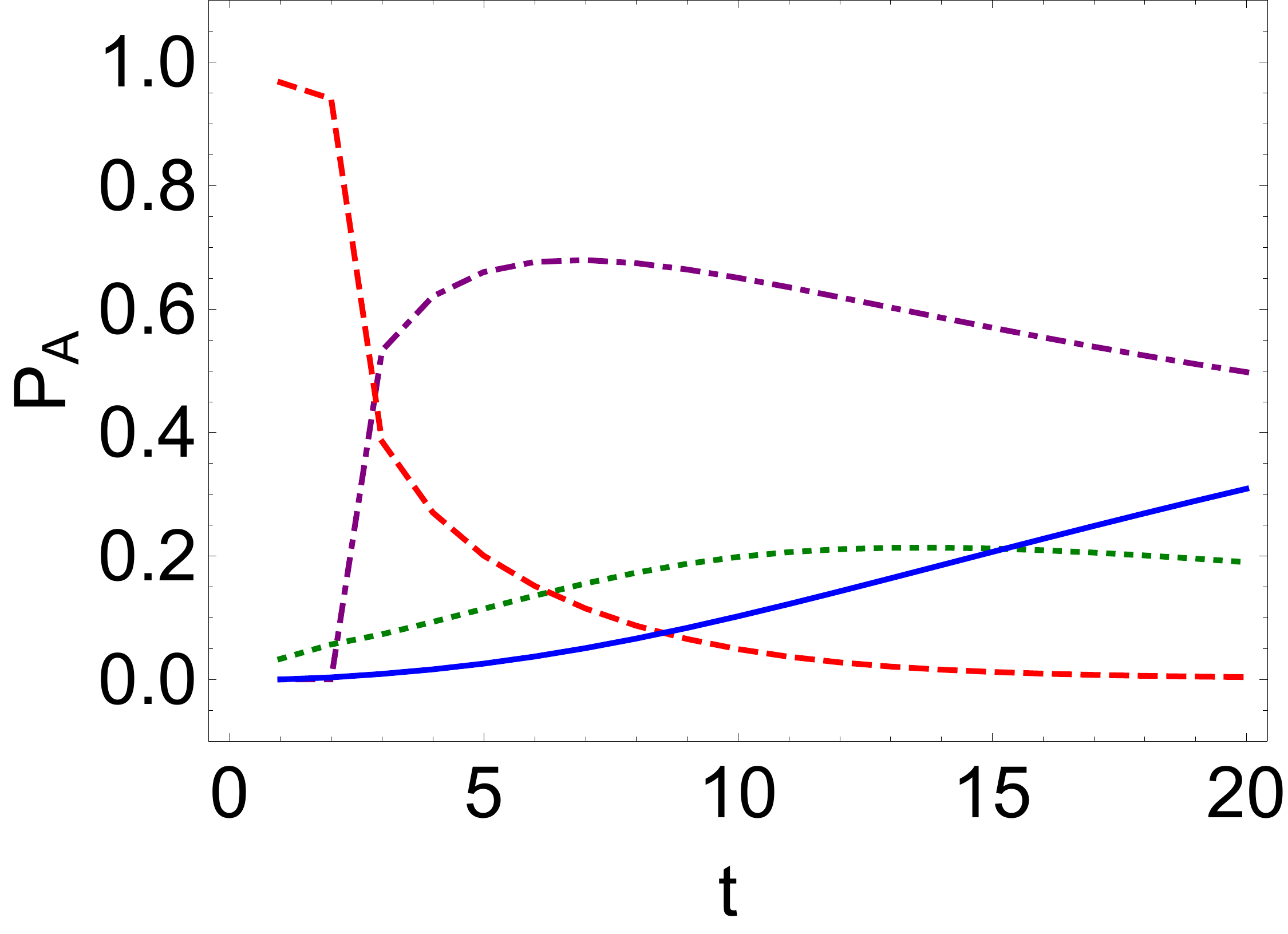}\quad
\includegraphics[width=3.5cm ]{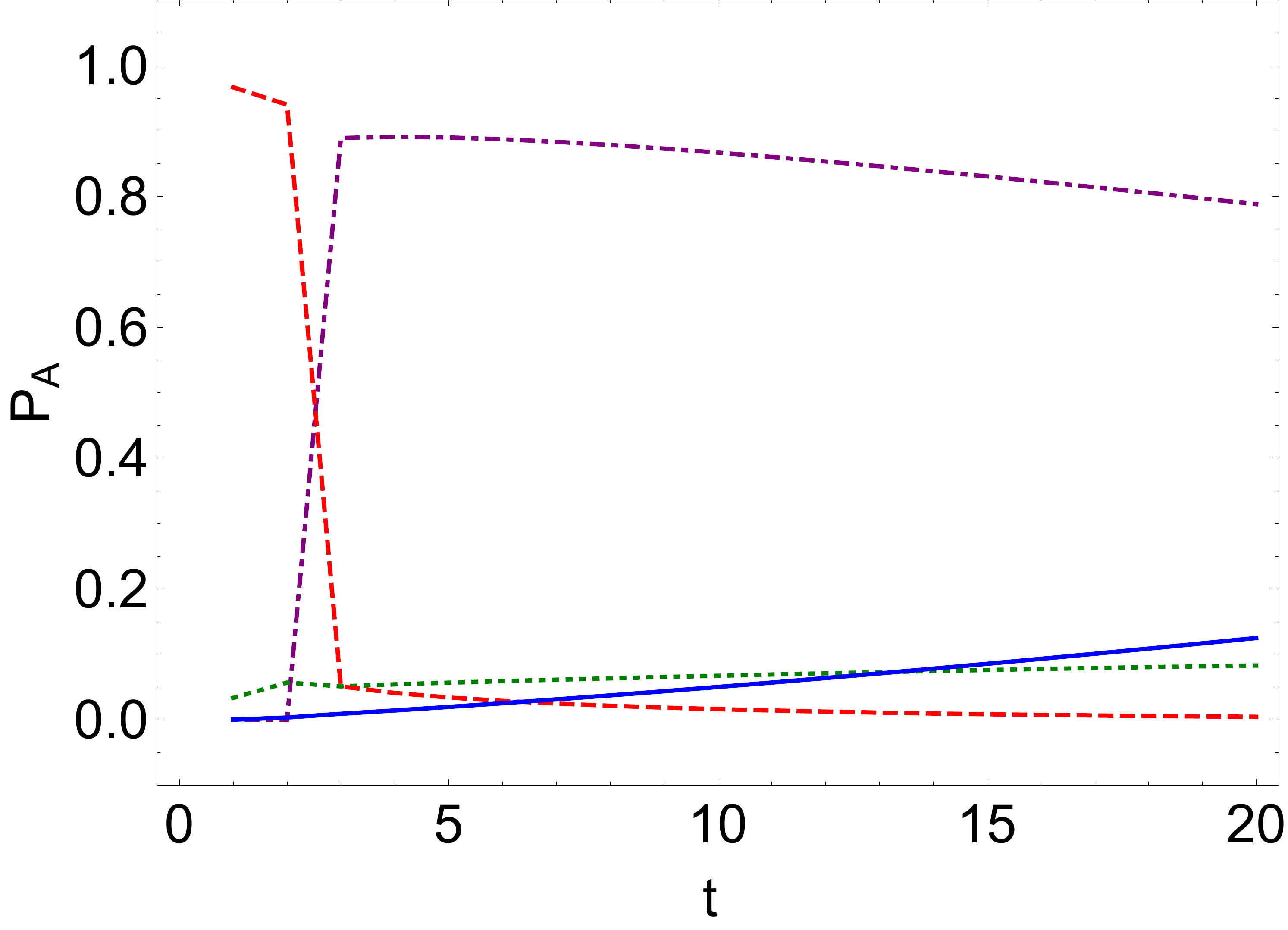}\\
\includegraphics[width=3.5cm ]{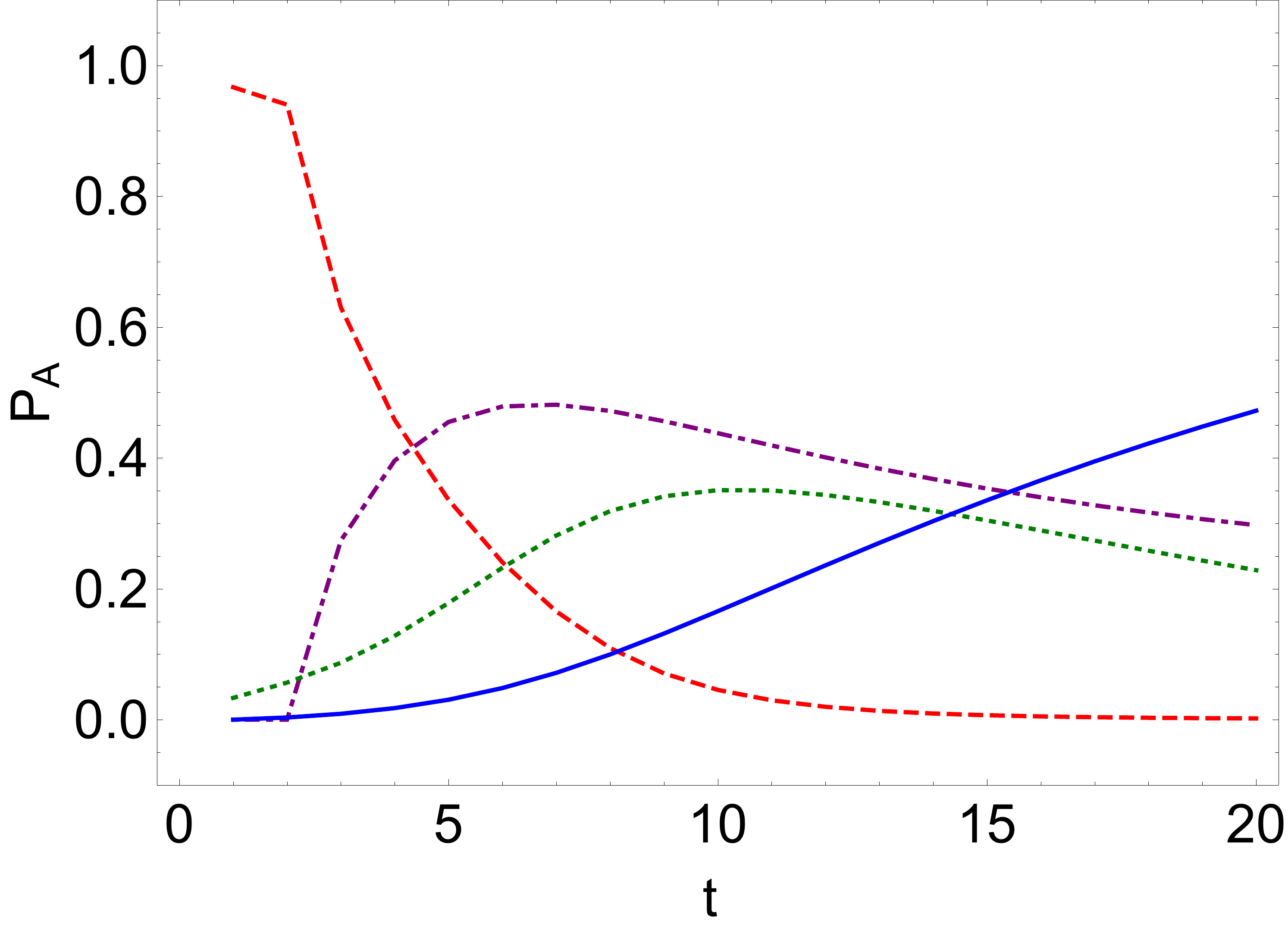}
\caption{Time dynamics of the probabilities 
in Equation \eqref{eq:prob} for the SIRV dynamics with only vaccination control
$\omega_k(t_I)$, the values used for the constant are $\alpha=0.2$, $\delta=0.1$, and $\gamma=0.02$. 
Different values for the constants were tested in \eqref{eq:cost}: top left $C_1=C_2=1$, 
top right $C_1=2C_2=2$, and bottom $2C_1=C_2=2$.
The Red dashed line is $P_S(t_i)$, 
the green dotted line represents $P_I(t_i)$, the purple dot dashed line represents $P_V(t_i)$, and the blue solid line represents $P_R(t_i)$.}
\label{fig:control-dyn}
\end{figure}
\begin{figure}
\centering
\includegraphics[width=3.5cm ]{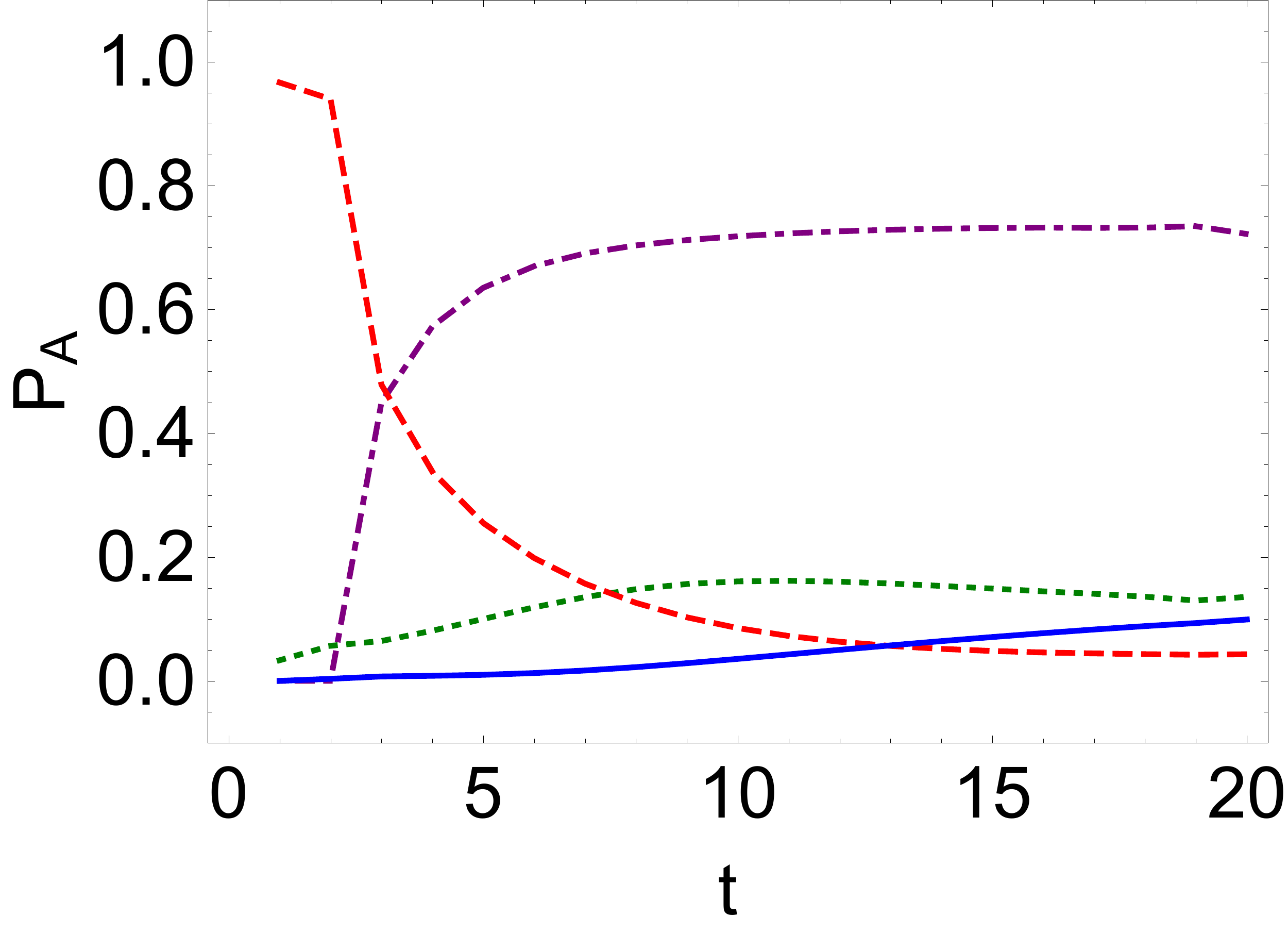}\quad
\includegraphics[width=3.5cm ]{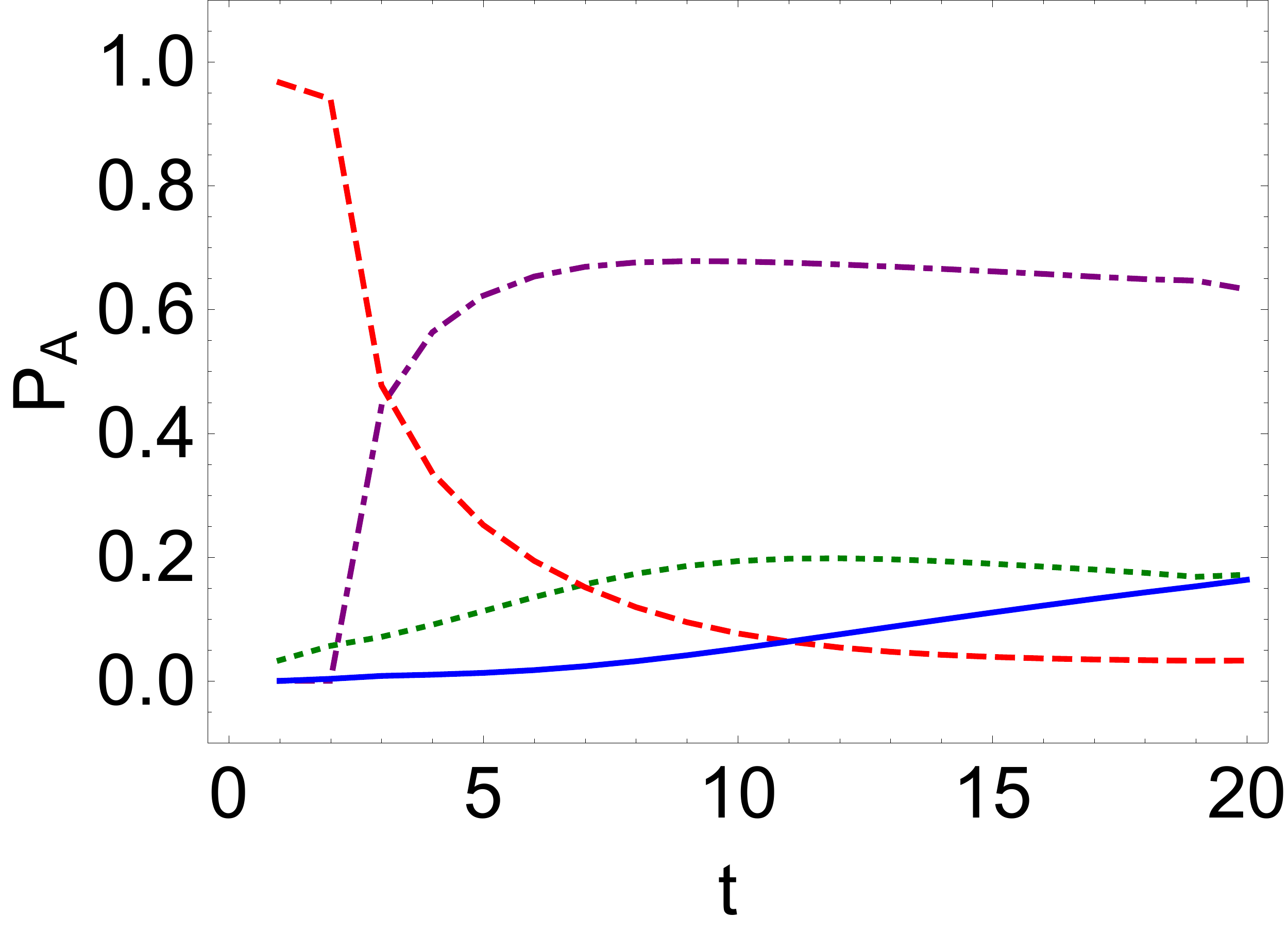}\\
\includegraphics[width=3.5cm ]{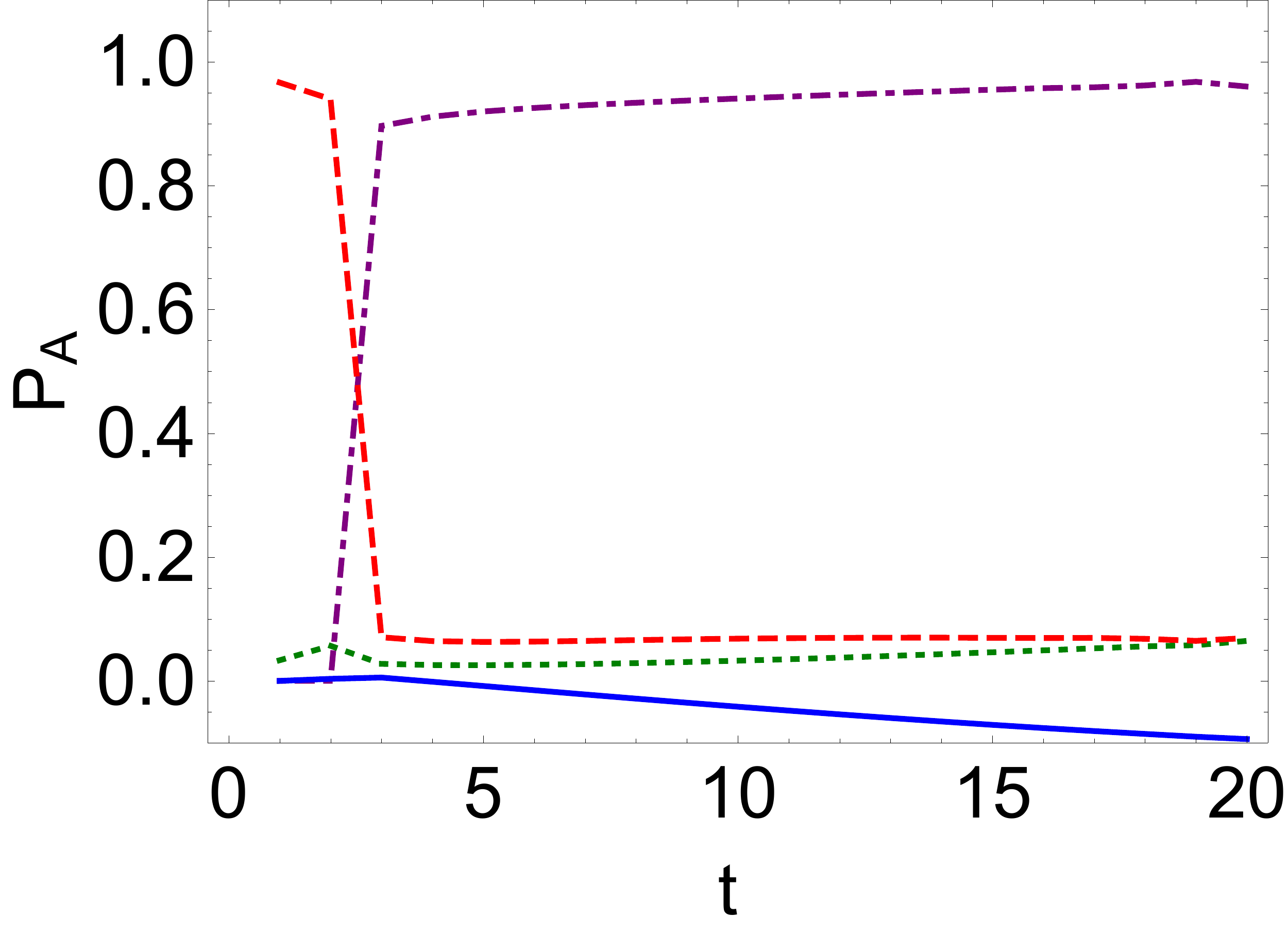}\quad
\includegraphics[width=3.5cm ]{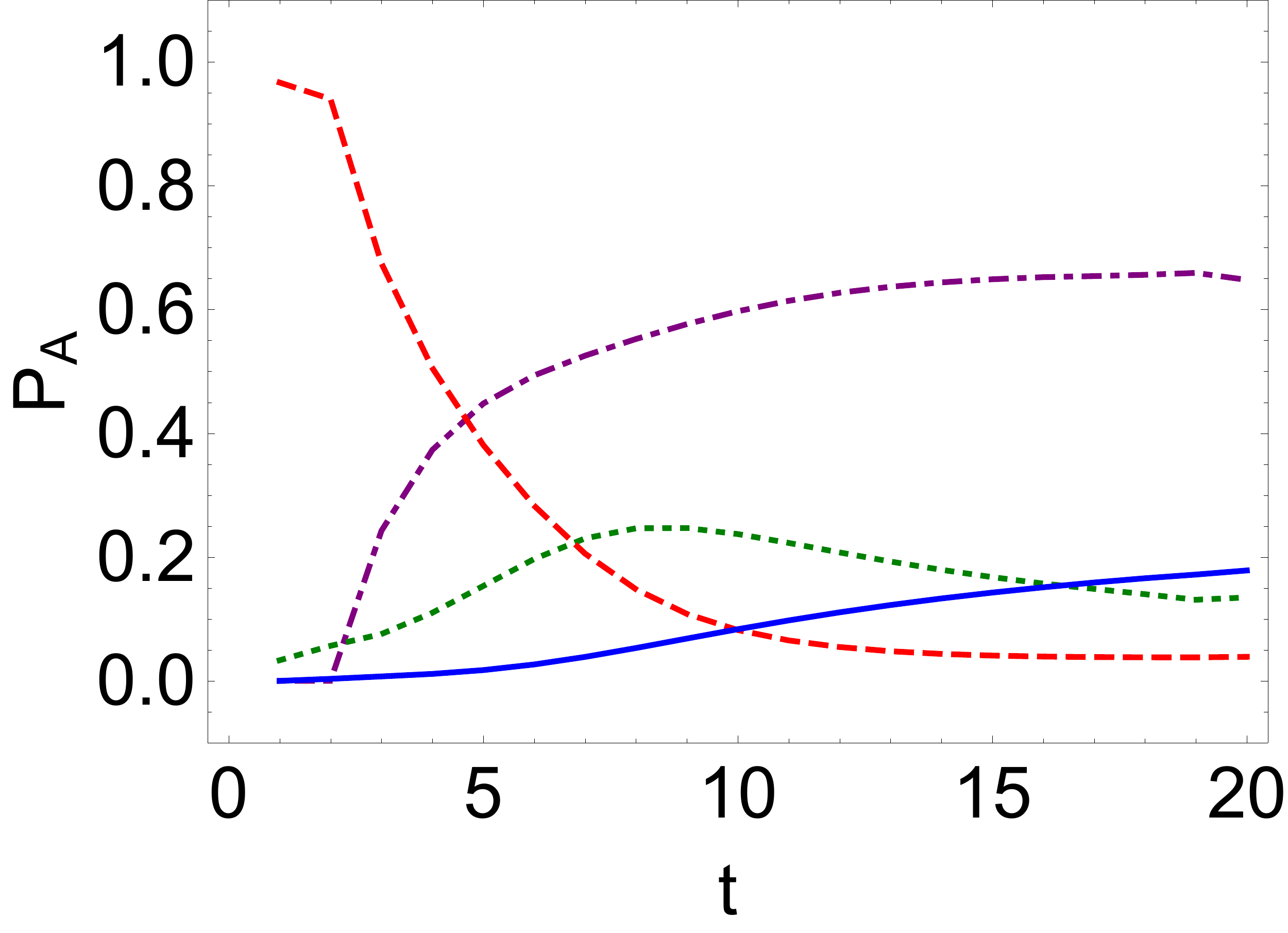}
\caption{Time dynamics of the probabilities 
in Equation \eqref{eq:prob} for the SIRV dynamics with both controls
$\omega_k(t_I)$ and $\tau_k(t_I)$; 
the values used for the constant are $\alpha=0.2$, $\delta=0.1$,$\theta=0.01$, and $\gamma=0.02$.
Different values for the constants were tested in \eqref{eq:cost-double}: top left $C_1=C_2=C_3=1$, 
top right $C_3=2C_1=2C_2=2$, bottom right $2C_1=C_2=2C_3=2$, and bottom left $C_1=2C_2=2C_3=2$.
The Red dashed line represents $P_S(t_i)$, 
the green dotted line represents $P_I(t_i)$, the purple dot dashed line represents $P_V(t_i)$, and the blue solid line represents $P_R(t_i)$.}
\label{fig:control-dyn-2}
\end{figure}
\begin{figure}
\centering
\includegraphics[width=6cm ]{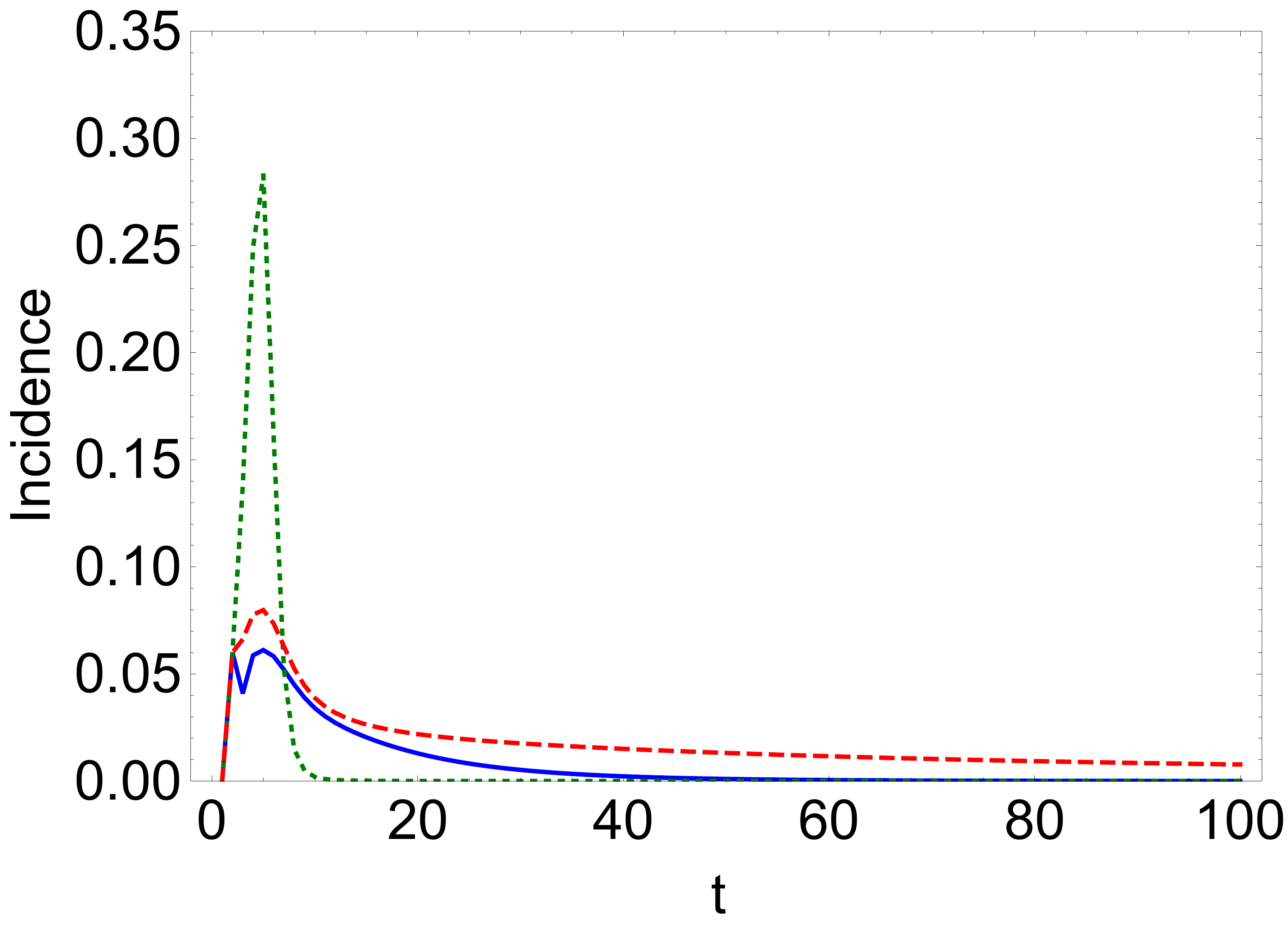}
\caption{The incidence in  three different cases, Green dotted line is the incidence of SIR model ,
Blue straight line is the case of single control, and Red dashed line represents the case with two controls. In this case the following constant are chosen $\alpha=0.2$, $\delta=0.1$, $\gamma=0.02$,
$\theta=0.01$ and $C_1=C_2=C_3=1$. }
\label{fig:incidence}
\end{figure}

To show some examples, 
a random complex network with 30 nodes is first chosen such that
the adjacency matrix is symmetric,
$a_{i,j}=a_{j,i},\quad\forall i,j=1,\dots 30$. 
In addition,
every node must have
at least one connection with another node, and the reference complex network
is shown in Figure \ref{fig:complex-net}.\\
Then, the initial condition is fixed by choosing random a node and setting it as $I_k(t_0)=1$.\\
The last choice made corresponds to the constants, and
they were set for the SIR model $\alpha=0.2$ and $\delta=0.1$.
The choice of these constants and the network was made to assure
that the system has critical behavior, in other words, for $t_{fin}\to \infty$,
$R_k(t_{fin})=1 \quad \forall k =1,\dots 30$, which is the so called
the susceptible-free equilibrium, in Figure \ref{fig:phase}, 
it is shown the phase diagram of SIR model in the reference network. 
In addition this behavior could be also seen in Figure \ref{fig:SIR-dyn}, and
these plots present the probability of a node to be susceptible, 
infected, or removed, and this quantity is defined as follows:
\begin{equation}
 P_A(t_i):=\frac{1}{N}\sum_{k=1}^NA_k(t_i),
 \label{eq:prob}
\end{equation}
with $A=R,I,S$ or $V$.\\
To obtain the optimal control, the standard forward-backward sweep
method is applied, and the procedure is the following:
\begin{itemize}
\item Step 1. Create a guess for all controls.
\item Step 2. Solve forward the dynamics of the system with the given control and initial conditions.
\item Step 3. Solve backward in time the adjoint equations.
\item Step 4. Evaluate new optimal conditions for controls. 
\item Step 5. Update the control with a weighted average between the old control 
 and the controls evaluated in Step 4.
\item Step 6. Check the convergence, if reached, stop; otherwise, go back to Step 2.
\end{itemize}
If $l$ is the index that represents the iteration number 
of the procedure stated, 
the criterion of convergence for the case with one control is the following equation:
\begin{equation}
\begin{array}{c}
 \forall k=1,\dots,N \qquad \forall ki=1\,\dots,t_{fin} \\
 \\
 \max\{\max\{\max\{|\omega_{k,l}(t_i) -\omega_{k,l+1}(t_i)|,|I_{k,l}(t_i)-I_{k,l+1}(t_i)| \}\\
 , |S_{k,l}(t_i)-S_{k,l+1}(t_i)| \}  ,|V_{k,l}(t_i)-V_{k,l+1}(t_i)|  \}<\rho,\\
\end{array}
 \end{equation}
in contrast, for the case with double control, the stop criterion is given by the following equation:
\begin{equation}
\begin{array}{c}
 \forall k=1,\dots,N \qquad \forall ki=1\,\dots,t_{fin} \\
 \\
 \max\{\max\{\max\{\max\{|\omega_{k,l}(t_i) -\omega_{k,l+1}(t_i)|,\\
 |I_{k,l}(t_i)-I_{k,l+1}(t_i)| \},
 |S_{k,l}(t_i)-S_{k,l+1}(t_i)| \} , \\
 |V_{k,l}(t_i)-V_{k,l+1}(t_i)|  \},
 \tau_{k,l}(t_i) -\tau_{k,l+1}(t_i)| \}<\rho,\\
 \end{array}
\end{equation}
in both cases, $\rho=10^{-4}$.
This criterion means that the
process is stopped only when the maximum distance between states and controls of
between two consecutive iterations is less than $\rho$.\\
Figure \ref{fig:control-dyn} shows the optimal 
solution of the SIRV problem with one control
for different
choices of the values of the constants in Equation \eqref{eq:cost}.
In contrast, Figure \ref{fig:control-dyn-2}
shows the optimal solution of the SIRV problem with two controls
for different
choices of the values of the constants in Equation \eqref{eq:cost}.
The optimal solution could indicate
a probabilistic vaccination
of a node, since $0\leq\omega_k(t_i)\leq 1$.
This fact is not a problem if a node represents
a metapopulation, but if a node represents a single individual, a partial vaccination 
is not understandable (and maybe ethically problematic). Attempts to solve this problem include
using a binary control, but in this case, the hypothesis of a piecewise continuous 
cost fails, and the optimal solution  could not be obtained .\\
Eventually to quantify the effects of the controls on the system
the incidence is presented, 
we compare the time evolution of incidence in the three models taken in account, results are
show in Figure \ref{fig:incidence}, in this figure it is possible to notice the effects of optimal control in reducing the 
but that in the case of double control the optimal solution for $C_1=C_2=C_3$ 
is to leave a very small probability that 
a nodes is infected so the infectious disease become endemic in that complex network, 
the average probability of a node to be infected by an another node after
100 time steps is very low 
$\sim 7\cdot 10^{-3}$ and this could not be true if the constants in the cost
are estimated in an another way,
however this behavior is
due to the fact that in the double control model taken in account 
a vaccinated individual could become susceptible again,
this feature gives ``supplies`` to the infected population as 
in the susceptible-infected-susceptible (SIS) model. 
Another fundamental contribution to this behavior is due to the fact the cost is quadratic in controls,
and the controls are bounded between $0$ and $1$ so, distribuiting
the action
in more steps is "cheaper" than doing the same action in a lesser number of steps.\\
Another quantity common in epidemiology is the
force of infection, defined in this case as $F(t_i)=\sum_{k=1}^N\alpha I_k(t_i)$,
again it is clear the effects of controls in reducing the ``size" of the infection,
and again the double control model show an endemicity of the disease.\\

\begin{figure}
\centering
\includegraphics[width=6cm ]{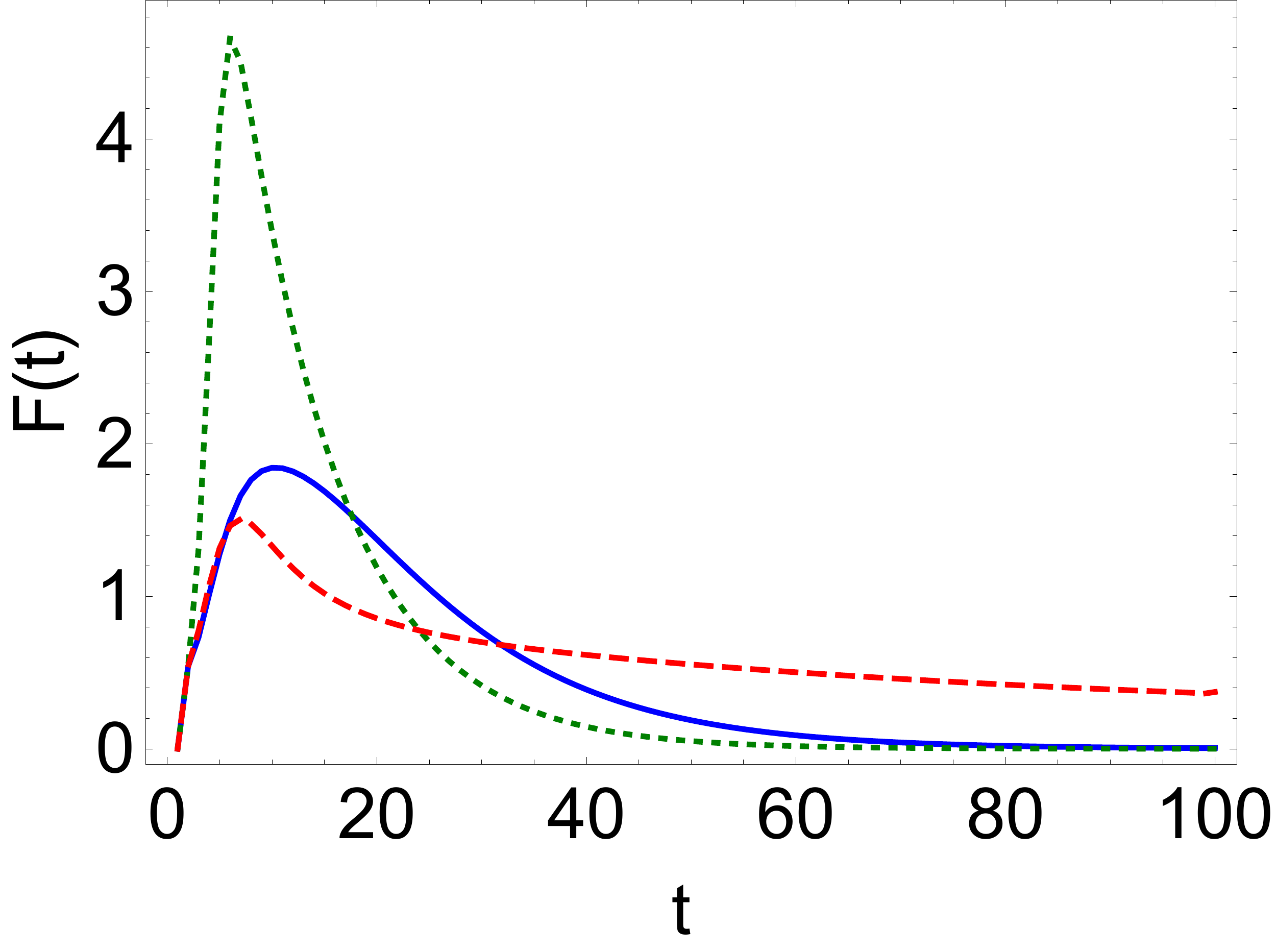}
\caption{The force of infection,
$F(t_i)=\sum_{k=1}^{N}\alpha I_k(t_i)$, of the three different cases taken in account, Gen dotted line $F_{t_i}$
of SIR model ,
Blue straight line is the case of single control, and Red dashed line  represents the case with two controls. In this case the following constant are chosen $\alpha=0.2$, $\delta=0.1$, $\gamma=0.02$,
$\theta=0.01$ and $C_1=C_2=C_3=1$. }
\label{fig:forc}
\end{figure}
\section{conclusion}
  Summarizing, the time discrete-time equations for the dynamics of the
  probability of individual nodes in the SIR model have been derived. In addition, 
  this method outperforms HMF and MC simulation \cite{sergi}, 
  since the proposed method requires small computational 
  effort and can accommodate probabilistic  (with some degree of uncertain)
  initial conditions that are common in epidemiology studies \cite{cit-prob-ci, surveliance-epi}.\\
  In addition, the framework for the SIRV was generalized, 
  and the optimal control in two different cases was then studied,
  considering only a vaccination strategy
  (transition from susceptible to vaccinated) and a double control 
  (transition from susceptible to vaccinated
  and a transition from infectious to vaccinated), 
  this study presents the set of equations needed
  to obtain the optimal control strategy given a cost, a network, and an initial probability
  distribution of infectious, susceptible, removed, and vaccinated.\\
  This kind of simulation could give an insight of how to balance cost 
 (e.g drugs production),
 epidemics constants (failure probability and duration of the vaccination),
 and distribution of health care
 in order to eradicate (or minimize) a endemic disease.\\ 
  Further works are required to overcome the problem of a probabilistic
  treatment. This method will also be applied to studying an infectious plant disease.
  Since in this case the topology of the complex network is known 
  and fixed in time, there is an interesting development of remote supervision  \cite{prob-tomato-1, cit-prob-ci, surveliance-epi-stem-root}
  of the disease, and the only concern in this case is that the SIRV dynamics could be not appropriate \cite{book-plant}
  for simulating a plant disease, so a tailored dynamic could be required.\\


  \end{document}